\documentclass[prb,preprint,floatfix,superscriptaddress]{revtex4-2}

\usepackage{amsmath,amssymb}
\usepackage{siunitx}
\usepackage{float}
\usepackage{graphicx}
\usepackage{pgf}
\usepackage{tikz}
\usepackage[colorlinks=true, citecolor=blue, linkcolor=blue, urlcolor=blue, pdfborder={0 0 0}]{hyperref}
\usepackage[capitalise]{cleveref}
\usepackage{subcaption}
\usepackage{blkarray}
\usepackage{filecontents}

\newcommand{\ket}[1]{|#1\rangle}
\newcommand{\creat}[1]{\hat{c}^{\dagger}_{#1}}
\newcommand{\annil}[1]{\hat{c}^{}_{#1}}
\newcommand{\creatcg}[1]{\hat{a}^{\dagger}_{#1}}
\newcommand{\annilcg}[1]{\hat{a}^{}_{#1}}

\newcommand{\dd}[2]{\frac{\mathrm{d}{#1}}{\mathrm{d}{#2}}}

\newcommand{\expe}{\mathrm{e}}
\newcommand{\compi}{\mathrm{i}}

\newcommand{\orcidicon}[1]{\href{https://orcid.org/#1}{\includegraphics[height=\fontcharht\font`\B]{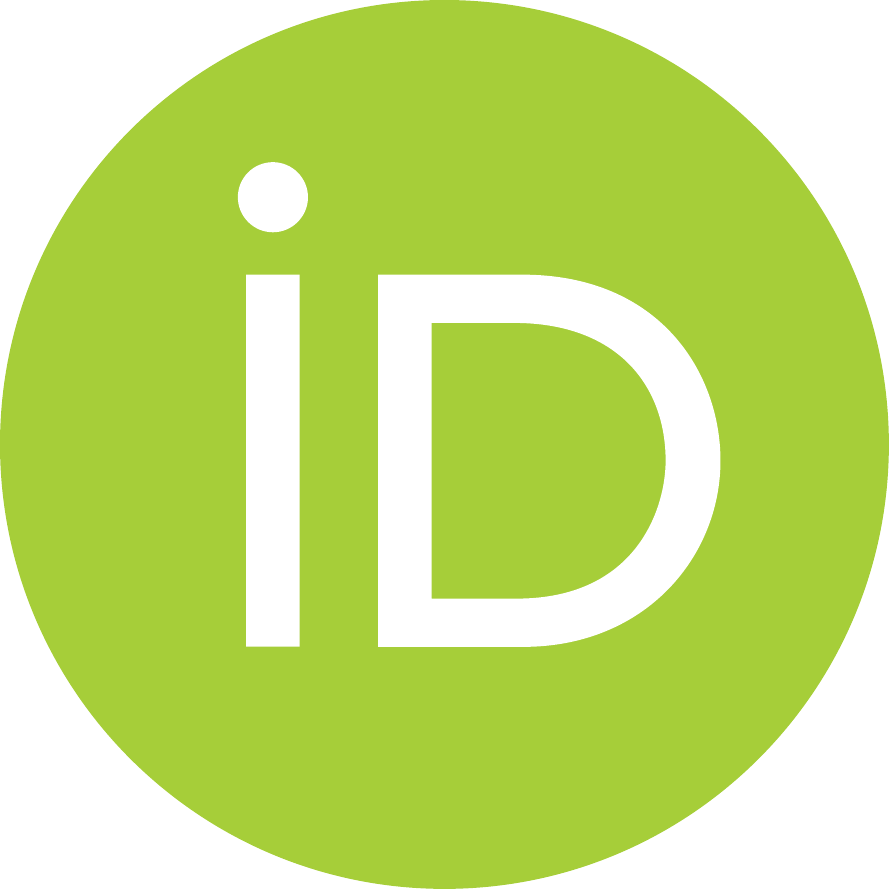}}}

\newlength{\myfigwidth}

\begin{document}

\title{Thermally Driven Polaron Transport in Conjugated Polymers}

\author{Laszlo Berencei\,\orcidicon{0000-0003-1609-4139}}
\email{laszlo.berencei@chem.ox.ac.uk}
\affiliation{Department of Chemistry, Physical and Theoretical Chemistry Laboratory, University of Oxford, Oxford, OX1 3QZ, United Kingdom}
\affiliation{Balliol College, University of Oxford, Oxford, OX1 3BJ, United Kingdom}
\author{William Barford\,\orcidicon{0000-0002-7223-686X}}
\email{william.barford@chem.ox.ac.uk}
\affiliation{Department of Chemistry, Physical and Theoretical Chemistry Laboratory, University of Oxford, Oxford, OX1 3QZ, United Kingdom}
\affiliation{Balliol College, University of Oxford, Oxford, OX1 3BJ, United Kingdom}
\author{Stephen R Clark\,\orcidicon{0000-0002-2072-7499}}
\affiliation{H.\ H.\ Wills Physics Laboratory, University of Bristol, Bristol, BS8 1TL, United Kingdom}

\begin{abstract}
We present a hybrid quantum-classical simulation of charge-polaron transport in conjugated polymers. The charge, which couples to the angular rotations of the monomers, is modeled via the time-dependent Schr\"odinger equation, while the monomers are treated classically via the Ehrenfest equations of motion. In addition, the system is thermalized by assuming that the monomers are subject to Brownian fluctuations modeled by the Langevin equation.
Charge coupling to the monomer rotations localizes the particle into a Landau polaron, while the thermal fluctuations of the monomers causes polaron dynamics.
The emergent low-energy scale of the model is the polaron reorganization energy, $E_r$, and thus $T_r = E_r/k_B$ is a convenient scale for the low-temperature dynamics.
We investigate two types of dynamics -- both relevant for temperatures $T < T_r$. In the lower temperature regime the system remains in the same quasidiabatic state, corresponding to activationless polaron diffusion as the polaron crawls stochastically along the chain. As the temperature is raised, however, there is a cross-over to an additional activated transfer process which corresponds to hopping between diabatic states. We show that these processes exhibit Landau-Zener type dynamics.
We note that as our model is general, it equally applies to exciton-polaron (i.e., energy) transport in conjugated polymers, and to charge and exciton polaron transport in quasi one-dimensional molecular stacks.
\end{abstract}

\maketitle

\section{Introduction}

Owing to the potential that $\pi$-conjugated polymer systems present for cheap and easily producible photovoltaic and light emitting devices, charge and energy (i.e., Frenkel exciton) transport in these systems has been intensely investigated for over 30 years.\cite{Kohler15} Understanding intrachain charge and exciton dynamics is complicated, because these particles are self-trapped by both fast (i.e., vibrational) and slow (i.e., torsional) degrees of freedom. The slow degrees of freedom also self-localize the particle into a Landau polaron.
In addition, thermal fluctuations of the torsional degrees of freedom results in dynamical off-diagonal disorder which both Anderson localizes the particle and causes polaron dynamics. The interplay between localization and thermally induced dynamics is the subject of this work.\par

Prins \emph{et al.}\cite{Prins06a, Prins06b} studied charge transport in polymers with static torsional disorder via the time-dependent Schr\"odinger equation, showing that disorder causes initially ballistic transport to become diffusive. Hultell and Strafstr\"om\cite{Hultell07} also used the time-dependent Schr\"odinger equation to model charge mobility in polymer chains with static disorder, but in addition they introduced polaronic effects via bond\cite{Hultell07} and torsional relaxation.\cite{Hultell09}\par

The role of thermally driven torsional fluctuations on charge transport was studied by Albu and Yaron.\cite{Albu13a} They considered the adiabatic regime (by assuming that the charge remains in its ground state) and found polaronic, activationless, diffusive behavior. Fornari and Troisi\cite{Fornari14} used the Fermi Golden Rule formalism to investigate the role of both static and dynamic torsional fluctuations on charge transport, while also including polaronic effects via bond relaxation. This approach necessarily models nonadiabatic processes, with the authors predicting both short range and longer-range hops via more delocalized states.\par

Poole \emph{et al.}\cite{Poole16} made a link between Ref.\cite{Albu13a} and Ref.\cite{Fornari14} by extending the work of Ref.\cite{Albu13a} to include nonadiabatic processes. The charge dynamics were simulated on the assumption that at time $(t+\delta t)$ the new target state is the eigenstate of $\hat{H}(t+\delta t)$ with the largest overlap with the previous target state at time $t$. However, owing to `trivial' crossings, this assumption was shown by Lee and Willard\cite{Lee19} to be potentially problematic for nonadiabatic transport.\par

Tozer and Barford\cite{Tozer15} used the approach of Poole \emph{et al.}\cite{Poole16} to model intrachain exciton (i.e., energy) transport in poly(para-phenylene) chains. Again, activationless exciton-polaron diffusion was predicted at low temperatures and activated exciton-polaron hopping at higher temperatures. A more sophisticated simulation of exciton motion in poly(p-phenylene vinylene) and oligothiophenes chains was performed by Burghardt and co-workers\cite{Binder20a, Binder20b, Binder20c,Hegger20} where high-frequency C-C bond stretches were also included, the solvent was modeled by a set of harmonic oscillators with an Ohmic spectral density, and the system was evolved via the multilayer-multiconfiguration time-dependent Hartree (MCTDH) method. Their results, however, are in quantitative agreement with those of Tozer and Barford in the `low-temperature' limit, namely activationless exciton diffusion with exciton diffusion coefficients close to experimental values. They did not observe thermally activated Marcus-like processes.\par

In this work we simulate both charge and energy transport in one-dimensional (1D) systems using the same generalized model. This model describes particle (i.e., charge or Frenkel exciton) delocalization along a 1D chain of lattice sites, where a site can represent a monomer in a $\pi$-conjugated polymer or a molecule in a 1D molecular stack. The particle couples to local harmonic oscillators via a change in the bond transfer integral caused by the relative displacement of the oscillators across the bond. The harmonic oscillators can model linear site displacements, in which case their relative displacement corresponds to a change of bond length or intermolecular spacing. Alternatively, as will be the example in this paper, the oscillators can model angular rotations of monomers, in which case their relative displacement corresponds to a change of the dihedral (or torsional) angle. Crucially, unlike earlier work,\cite{Poole16,Tozer15} the full charge and exciton dynamics are determined by the time-dependent Schr\"odinger equation, thus removing ambiguities at energy-level crossings.\cite{Lee19}
The dynamics of the classical oscillators, however, are described by the Ehrenfest equations of motion. We model system-bath interactions by coupling the harmonic oscillators to the Brownian fluctuations of the bath, specifically via the Langevin equation.

The decision to model the harmonic oscillators and the bath classically, via the Ehrenfest and Langevin equations, respectively, is an expediency motivated by the necessity of simulating large systems (over 50 monomer `sites'), for long times (over $50$ ps), and to perform extensive ensemble averaging. However, the limitations of the Ehrenfest approximation are well-documented\cite{Horsfield06} and we discuss these in relation to our results in \cref{sec:conclusions}.

This problem exhibits rich physical behavior, which we discuss qualitatively here before quantifying these remarks later in the paper.
As stated above, from now on we assume that the charge or exciton couples to torsional fluctuations via angular rotations of the monomers in a conjugated polymer.
In the absence of system-bath interactions, i.e., at zero-temperature, a classical treatment of the harmonic oscillators implies that their coupling to the particle self-traps the particle into a Landau polaron. That is, the particle `digs a hole' for itself via local, static displacements of the oscillators and becomes self-localized.\footnote{This only true in the adiabatic limit of slow oscillators\cite{Tozer14}.} Here, we consider the `large-polaron' limit where the spatial extent of the polaron extends over many monomers.\par

As the temperature is raised, small thermally-induced torsional fluctuations cause the polaron to `crawl' diffusively along the chain, remaining in the same quasidiabatic state with no diabatic energy level crossings. In this activationless limit the diffusion coefficient is proportional to temperature.
As the temperature is raised still further thermally activated processes become more important. In particular, larger torsional fluctuations cause the energy of the localized particle to match those of neighboring localized states leading to diabatic energy level crossings and Landau-Zener type dynamics.
This manifests itself as Marcus-type hopping between quasidiabatic states. Such behavior has been rigorously investigated via instanton theory using model two-level systems\cite{Garg1985,Takahashi2017}.
\par

At still higher temperatures the polaron becomes unbound and the thermal fluctuations in angular displacements cause temporally and spatially varying off-diagonal disorder that localizes the particle into Anderson polarons, i.e., single particle states localized by disorder.\cite{Anderson58a,Malyshev01a,Malyshev01b} The transport in this limit is quasiband-like.\cite{Troisi06,Wang15}\par

We simulate the first two of these processes by the generalized model introduced in \cref{subsec:Hamiltonian} with the methodologies described in \cref{subsec:QD,subsec:Ehren}. We present our results in \cref{sec:results} and conclude in \cref{sec:conclusions}. In this work we focus on charge-polaron dynamics, although our results equally well apply to excitons.

\section{Model and Methodology}\label{sec:methods}

\subsection{Model Hamiltonian}\label{subsec:Hamiltonian}

As a concrete example of polaron transport in quasi-1D systems, we consider charge motion in a conjugated polymer, i.e., poly(para-phenylene) (PPP), where the phenyl rings undergo simple harmonic motion. We assume that the oscillators are independent and execute small angular displacements, $\Delta{\phi}$, from equilibrium.
Then, as shown in \hyperref[App:1]{the Appendix}, by introducing dimensionless variables for the conjugate variables $\phi$ and the monomer angular momentum, $L$, i.e.,
\begin{equation}
  \tilde{\phi} = (K/\hbar \omega)^{1/2}\phi
\end{equation}
and
\begin{equation}
  \tilde{L} =(\omega/\hbar K)^{1/2} L,
\end{equation}
 the coarse-grained Hamiltonian for the system is
\begin{align}\label{eq:Ham}
\nonumber \hat{H} = &\sum_{n=1}^{N} \left[J_{n}^0
-\hbar\omega  A_n (\Delta \tilde{\phi}_{n+1} - \Delta \tilde{\phi}_{n})\right]\hat{T}_{n,n+1} \\
+ &\frac{\hbar \omega}{2}  \sum_{n=1}^{N} \left( (\Delta\tilde{\phi}_n)^2 + \tilde{L}_n^2 \right).
\end{align}
The `site' index, $n$, labels a monomer, $N$ is the number of monomers, $\omega = (K/I)^{1/2}$ is the rotational angular frequency, $K$ is the elastic force constant and $I$ is the moment of inertia of each monomer.
\begin{equation}
 \hat{T}_{n,n+1} = \left(\hat{a}^{\dagger}_{n+1}\hat{a}_{n} + \hat{a}^{\dagger}_{n}\hat{a}_{n+1}\right)
\end{equation}
is the bond-order operator, where $\hat{a}^{\dagger}_{n}$ ($\hat{a}_{n}$) creates (destroys) a charge (or Frenkel exciton) on monomer $n$.

\begin{figure}
    \centering
    \includegraphics[width=0.9\myfigwidth]{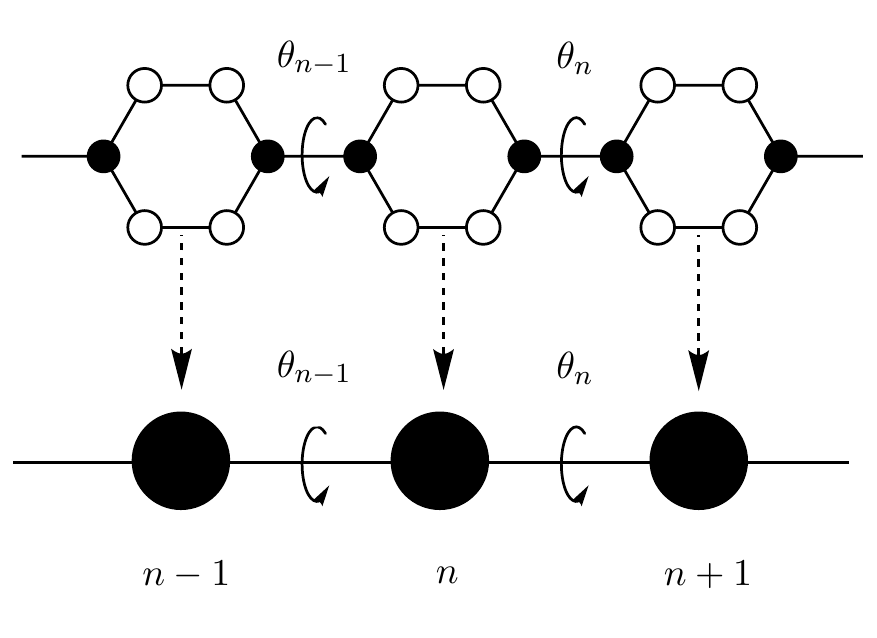}
\caption{\label{fig:1} The coarse-graining of PPP to a linear chain of `sites'. The $p_z$ atomic amplitudes on the phenyl ring are $1/\sqrt{3}$ for filled circles and $-1/\sqrt{12}$ for unfilled circles. $\theta_n = (\phi_{n+1}-\phi_n)$ is the torsional angle of the $n$th bond, where $\phi_n$ is the rotational angle of the $n$th ring with respect to the molecular axis.}
\end{figure}

$J_n^0$ is the transfer integral between monomers and $A_n$ is the dimensionless parameter that describes the coupling between the charge (or exciton) and the harmonic oscillators.
Evidently, the angular fluctuations of the monomers, $\Delta \phi_n$, cause torsional fluctuations (or bond rotations), $\Delta \theta_n = (\Delta \phi_{n+1} - \Delta \phi_n)$, which couple to the bond-order operator.
The last two terms in \cref{eq:Ham} are the monomer elastic and kinetic energies.
\Cref{eq:Ham} is derived for charges in \hyperref[App:1]{the Appendix} and in Ref.\cite{Barford18} for Frenkel excitons.
The mapping from the PPP atomic structure to a coarse-grained linear chain of `sites' is illustrated in \cref{fig:1}.

The charge transfer integral is
\begin{equation}
J_n^0 = \beta_{0} \cos \theta_n^0/3,
\end{equation}
where $\beta_{0}$ is the H\"uckel resonance integral for neighboring parallel $p_z$ orbitals, $\theta_n^0$ is the equilibrium torsional angle of bond $n$ in the neutral state (or ground state for excitons) and
\begin{equation}
A_n = \frac{\beta_{0} \sin  \theta_n^0}{3(\hbar \omega K)^{1/2}}.
\end{equation}
Corresponding expressions for Frenkel excitons are given in \cref{sec:conclusions}.

\subsection{Landau Polarons}\label{subsec:Landau}

The dimensionless torque on each monomer is
\begin{align}
\nonumber
\Gamma_n &= - \frac{\partial \langle \hat{H} /\hbar \omega\rangle}{\partial \Delta\tilde{\phi}_n}\\
\label{eq:torque} &= -\Delta\tilde{\phi}_n +\lambda_n,
\end{align}
where
\begin{equation}
\lambda_n = A_{n-1} \langle \hat{T}_{n-1,n} \rangle - A_{n} \langle \hat{T}_{n,n+1} \rangle
\end{equation}
is the torque exerted by the charge.
Setting $\Gamma_n = 0$ gives the equilibrium angular displacements in the charged state (or the excited state for excitons) as
\begin{equation}
\Delta\tilde{\phi}_n^{\textrm{eq}} = \lambda_n.
\end{equation}

For a polymer chain with staggered monomer displacements in the neutral state (or ground state for excitons), i.e., $\theta_{n+1} = - \theta_n$, the equilibrium displacement in the continuum limit satisfies\cite{Barford18}
\begin{equation}\label{Eq:10}
 \Delta\tilde{\phi}_n^{\textrm{eq}} \sim \rho_n \times (-1)^n,
\end{equation}
where $\rho_n$, the charge (exciton) density, is
\begin{equation}\label{Eq:13}
  \rho_n = \left(\frac{\kappa}{2}\right) \textrm{sech}^2 \kappa(n -n_0)
\end{equation}
and
\begin{equation}\label{Eq::11a}
\kappa = A^2 \hbar \omega/4J.
\end{equation}
The `large-polaron' limit is  $\kappa^{-1} \gg 1$, implying a sufficiently small electron-phonon coupling.
Thus, the particle is self-trapped and self-localized as a Landau polaron with a reorganization energy in the continuum limit,\cite{Holstein59a,Rashba82,Book}
\begin{gather}
\label{eq:E_b} E_r = \frac{J}{12}\left(\frac{\hbar \omega}{J}\right)^2 \left( \frac{A^2}{2} \right)^2.
\end{gather}
The polaron reorganization energy is the emergent low-energy scale from \cref{eq:Ham}, so $T_r = E_r/k_B$ serves as a convenient temperature scale for the low temperature dynamics.

We note that the classical, Landau polaron is rigorously valid in the limit that $\hbar \omega/J \rightarrow 0$, as then the oscillators respond infinitesimally slowly to the charge dynamics so that the polaron effective mass diverges\cite{Tozer14}. The center of the polaron, $n_0$, is determined by disorder and as we see later, if the disorder is dynamical $n_0$ varies with time implying polaron dynamics.

\subsection{Quantum Dynamics}\label{subsec:QD}

The time-dependent Schr\"odinger equation is solved for the charge (or exciton) via
\begin{equation}
  |\Psi(t+\delta t) \rangle = \exp(- \mathrm{i} \hat{H}\delta t/\hbar)\vert \Psi(t) \rangle
\end{equation}
using the TNT Library.\cite{Al_Assam_2017} The TNT Library is a computer package designed to simulate 1D quantum systems using the matrix product state formalism. Dynamics is described by the time-evolving block decimation (TEBD) scheme,\cite{Vidal2003,Vidal2004} which is coupled with a classical evolution of the angular displacements according to the Ehrenfest equations of motion. Periodic boundary conditions were implemented in the TNT library in this work, which is described in \hyperref[sec:supp]{the Supplemental Material} (see, also, Refs. \cite{Trotter1959,Suzuki1976,DeRaedt1983}, therein).

\subsection{Ehrenfest and Langevin Dynamics}\label{subsec:Ehren}

The classical angular displacements, $\{ \Delta \tilde{\phi}\}$, are subject to the Ehrenfest equations of motion,
\begin{equation}\label{eq:18}
\dd{\Delta\tilde{\phi}_n}{\tilde{t}} = \tilde{L}_n
\end{equation}
and
\begin{equation}\label{eq:19}
\dd{\tilde{L}_n}{\tilde{t}}  =  \Gamma_n  -\tilde{\gamma} \tilde{L}_n + R_n(t),
\end{equation}
where $\tilde{t} = \omega t$ and $\tilde{\gamma} = \gamma/\omega $ is a dimensionless friction coefficient.

To model the role of temperature we supplement the systematic torque on each monomer, $\Gamma_n$, with a dimensionless random torque, $R_{n}(t)$. For a polymer in solution, the origin of this torque is the Brownian fluctuations of the solvent molecules.
By the fluctuation-dissipation theorem, the random torques satisfy
\begin{equation}\label{fluc_diss}
\langle R_{n}(0)R_{m}(t) \rangle = 2\tilde{\gamma} k_{B}T\delta_{mn}\delta(\tilde{t})/\hbar\omega.
\end{equation}
In addition to being both spatially and temporally uncorrelated, these torques are also assumed to be independent of the angular velocity of the phenylene ring and independent of the systematic torque, $\Gamma_n$. The stochastic torques then possess a Gaussian distribution with a standard deviation of
\begin{equation}
\sigma_{R} = (2 \tilde{\gamma} k_{B}T/\hbar\omega\Delta \tilde{t})^{\frac{1}{2}}.
\end{equation}
In the absence of polaronic coupling, the Brownian fluctuations cause a random distribution of angular displacements, whose variance satisfies the principle of equipartition, i.e.,
\begin{equation}
 \langle  \Delta \tilde{\phi}^2 \rangle = \frac{k_B T}{\hbar \omega}.
\end{equation}
As may be seen from \cref{eq:Ham}, these fluctuations cause off-diagonal disorder in the particle hopping and for temperatures $T > T_r = E_r/k_B$ induce Anderson localization of the particle wave function.\par

We implement the Langevin dynamics using the algorithm described by Gr{\o}nbech-Jensen and Farago,\cite{Jensen2013} a position-Verlet integration scheme.

\subsection{Simulation Parameters}

The parameters used in the simulations of charge-polaron dynamics in PPP are listed in \cref{tab:params}. The simulations were performed on chains of $N=50$ sites (i.e., 50 coarse-grained monomers) with periodic boundary conditions and ensemble averaging over at least 25 different trajectories at each temperature. The dynamics in each instance are initiated in a localized polaron state, obtained via a Hellmann-Feynman (HF) minimization routine. This solver iteratively finds the equilibrium angular displacements via the torque expression in \cref{eq:torque} using the equilibrium condition of $\Gamma_{n}^{\mathrm{eq}} = 0$. The shape of this localized polaron state is shown in the $t=0$ snapshot in Fig.\ 2 and satisfies \cref{Eq:13}.
The polaron reorganization energy is obtained numerically via the HF iterator.
The simulation time for each trajectory is $\sim 7 \pm 3$ days.
\begin{table}[H]
\caption{\label{tab:params}Parameters\cite{Barford18,Berencei19} used in the simulations of charge-polaron dynamics in PPP.}
\begin{ruledtabular}
\begin{tabular}{lcc}
Roton energy & $\hbar\omega$ & $0.02\,\si{\electronvolt}$ \\
Charge transfer integral & $J$ & $-0.689\,\si{\electronvolt}$ \\
Charge-roton coupling constant & $A$ & $1.87$ \\
Ground state torsional angle & $\phi^{0}$ & $\pm 10^{\circ}$ \\
Dissipation coefficient & $ \gamma$ &  $10^{12}\,\si{\per\second}$ \\
Total simulation time &  & $300\,\tau$ \\
Rotational period & $\tau = 2\pi/\omega$  & $0.2\,\si{\pico\second}$ \\
Charge-polaron reorganization temperature & $T_r = E_r/k_B$  & $\sim 400 \,\si{\kelvin}$ \\
Quantum dynamics time step &  & $1.6\times 10^{-5} \, \tau$ \\
Classical dynamics time step &  & $1.6\times 10^{-3} \, \tau$
\end{tabular}
\end{ruledtabular}
\end{table}

\section{Results}\label{sec:results}

\subsection{Low temperature polaronic crawling}

\begin{figure*}[h!]
    \centering
    \includegraphics[width=1.0\linewidth]{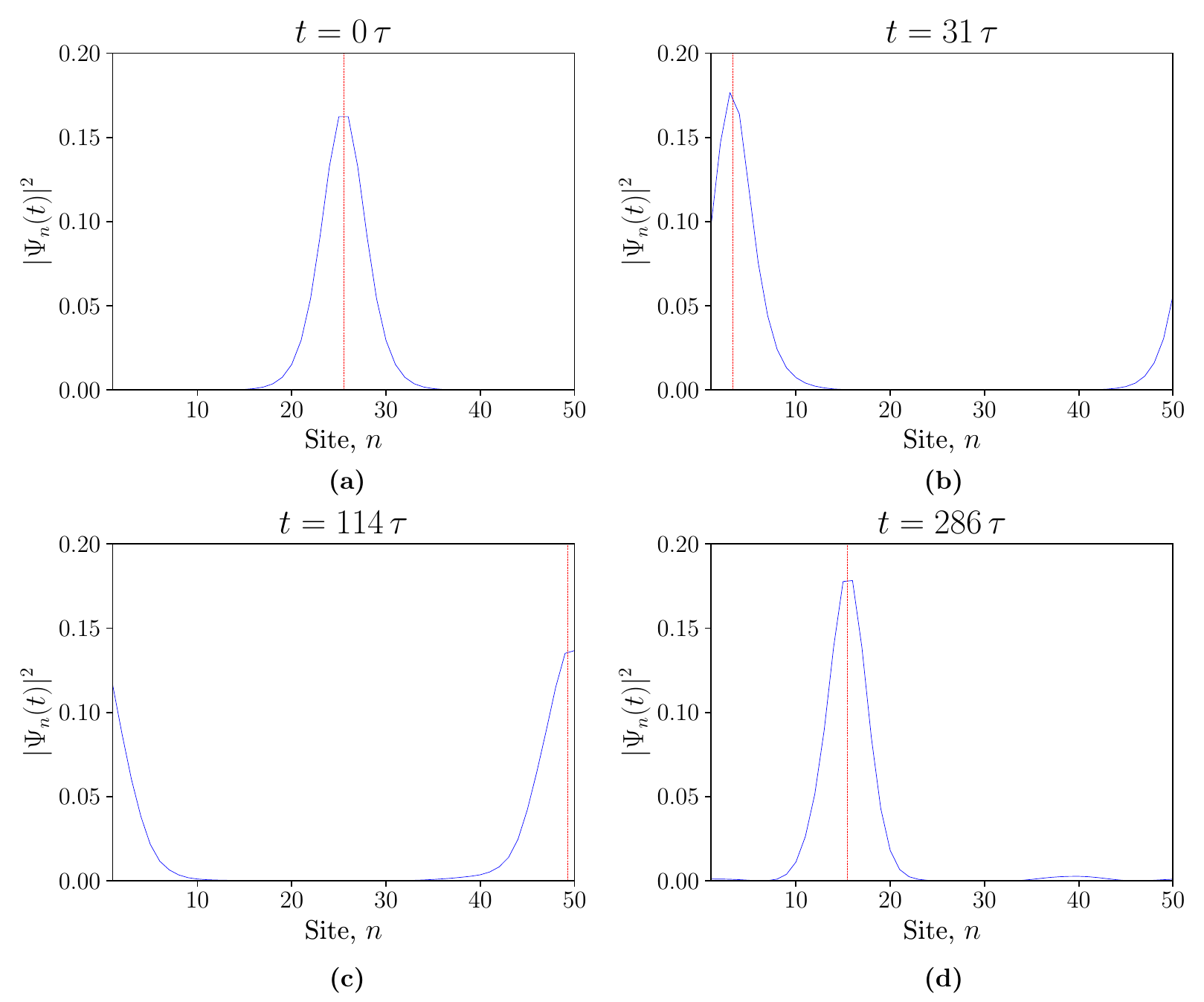}
    \label{fig:2}
    \caption{Polaron density as a function of time at $T = 0.024\, T_r$. $\tau$ is the monomer rotational period. The center of mass is shown as the red vertical line.}
\end{figure*}

\begin{figure}[h!]
    \centering
    \includegraphics[width=0.9\myfigwidth]{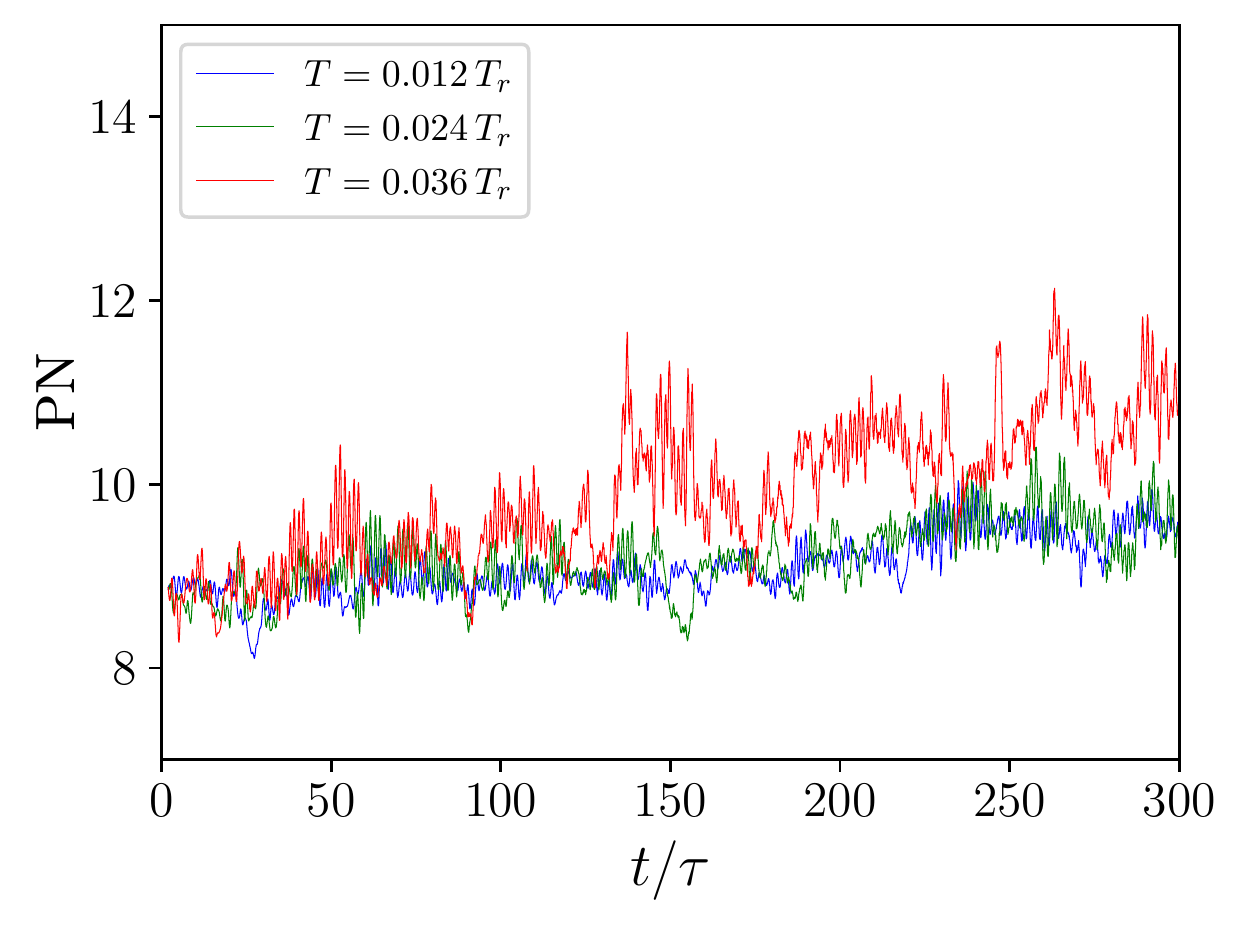}
    \caption{\label{fig:3} Participation number (i.e., localization length) of the evolving state, $|\Psi(t)\rangle$, as a function of time in the lowest temperature regime, $T \ll T_r$. The curves are smoothed for better legibility.}
\end{figure}

\begin{figure}[h!]
    \centering
    \includegraphics[width=0.9\myfigwidth]{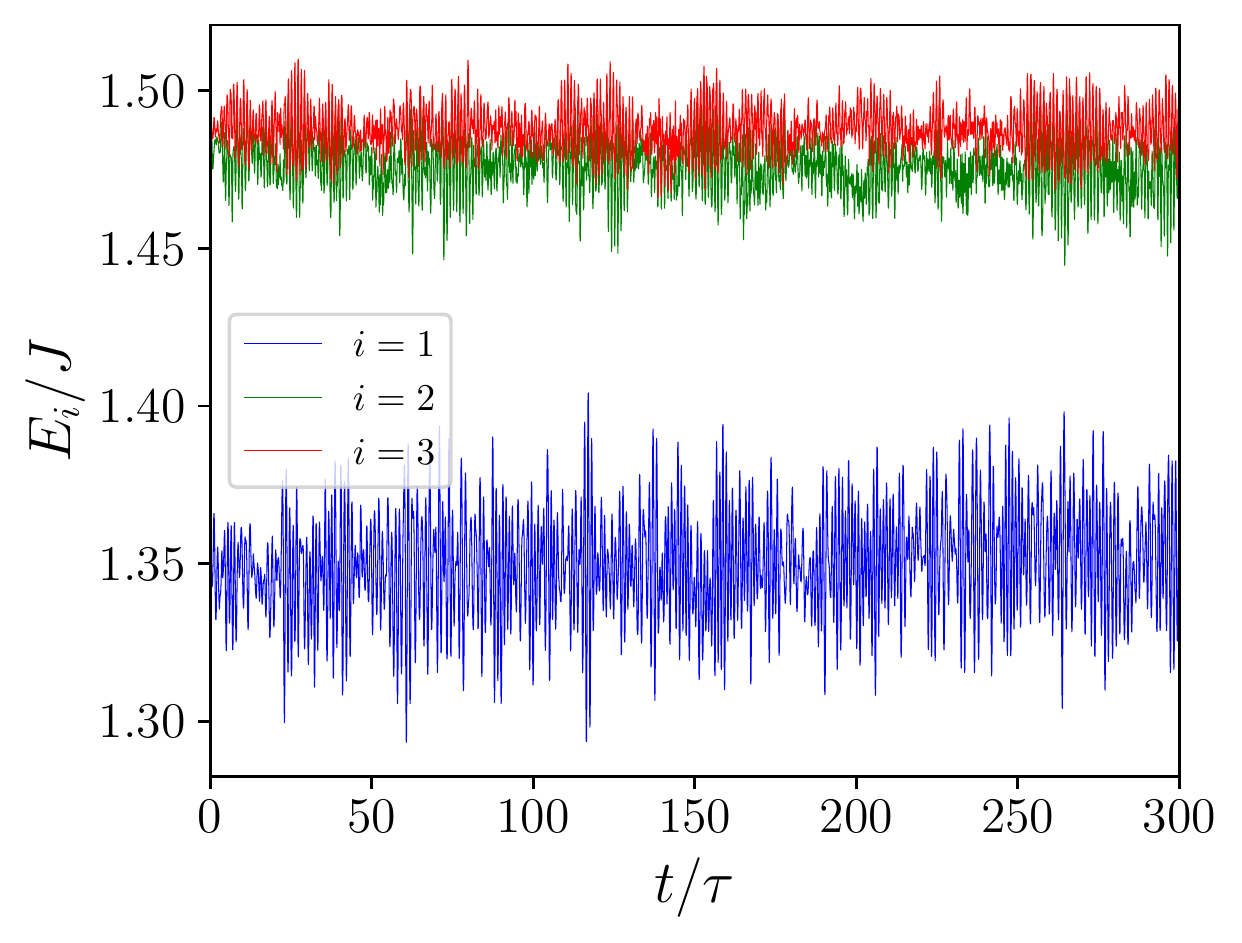}
    \caption{\label{fig:4} Energies of the lowest instantaneous eigenstates as a function of time at $T = 0.024\, T_r$. }
\end{figure}

\begin{figure}[h!]
    \centering
    \includegraphics[width=0.9\myfigwidth]{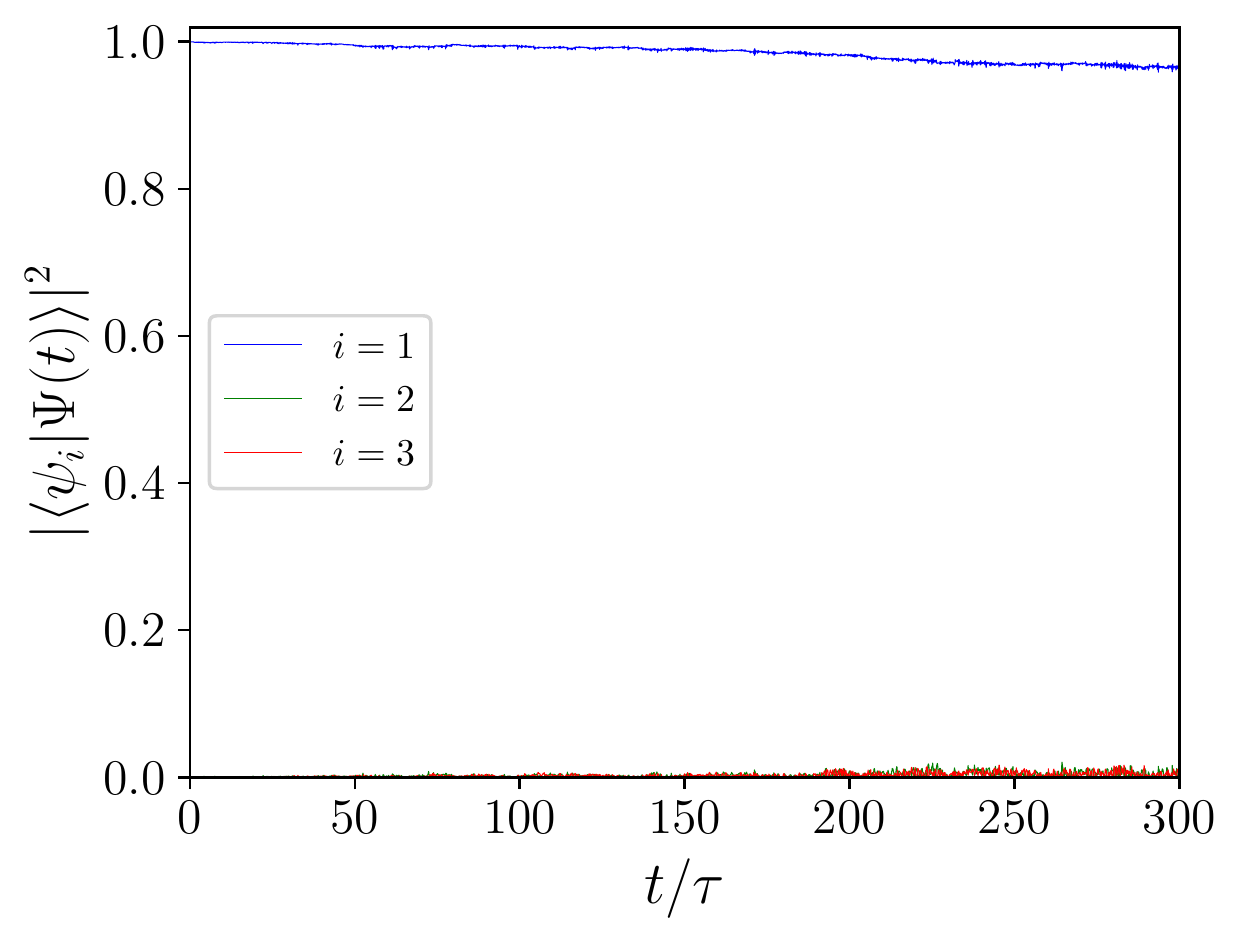}
    \caption{\label{fig:5} Probabilities that the evolving state, $|\Psi(t)\rangle$, occupies an instantaneous adiabatic state, $|\psi_i\rangle$, as a function of time at $T = 0.024\,T_r$.}
\end{figure}

The dynamics in the low temperature regime, defined as $T \ll T_r$, follows a quasidiabatic evolution in the lowest energy diabatic state. This leads to a `crawling' displacement of the polaron along the polymer chain, and therefore to a net migration of charge (or energy). The displacement of the polaron is determined by tracking the center of mass, i.e., $\bar{n}(t) = \sum_{n} n\rho_{n}(t)$, over time. Snapshots of the evolution of the polaron density as a function of time, illustrated in \cref{fig:2}, show that at low temperatures the polaron moves along the full length of the chain and freely across the periodic boundaries, retaining its shape at all times. The localization length, defined by the participation number, PN, as
\begin{gather}
\textrm{PN} = \frac{1}{\sum_{n} \left\vert \Psi_{n} \right\vert^{4}},
\end{gather}
where $\Psi_n$ denotes the evolving wave function amplitude on site $n$, remains small throughout the simulations, as shown in \cref{fig:3}. The diabatic nature of the evolution is also demonstrated by the absence of energy crossovers between the lowest instantaneous eigenstates of the Hamiltonian (shown in \cref{fig:4}), and by considering the overlap of the evolving wave function with these adiabatic eigenstates, shown in \cref{fig:5}.

\begin{figure}[h!]
    \centering
    \includegraphics[width=0.9\myfigwidth]{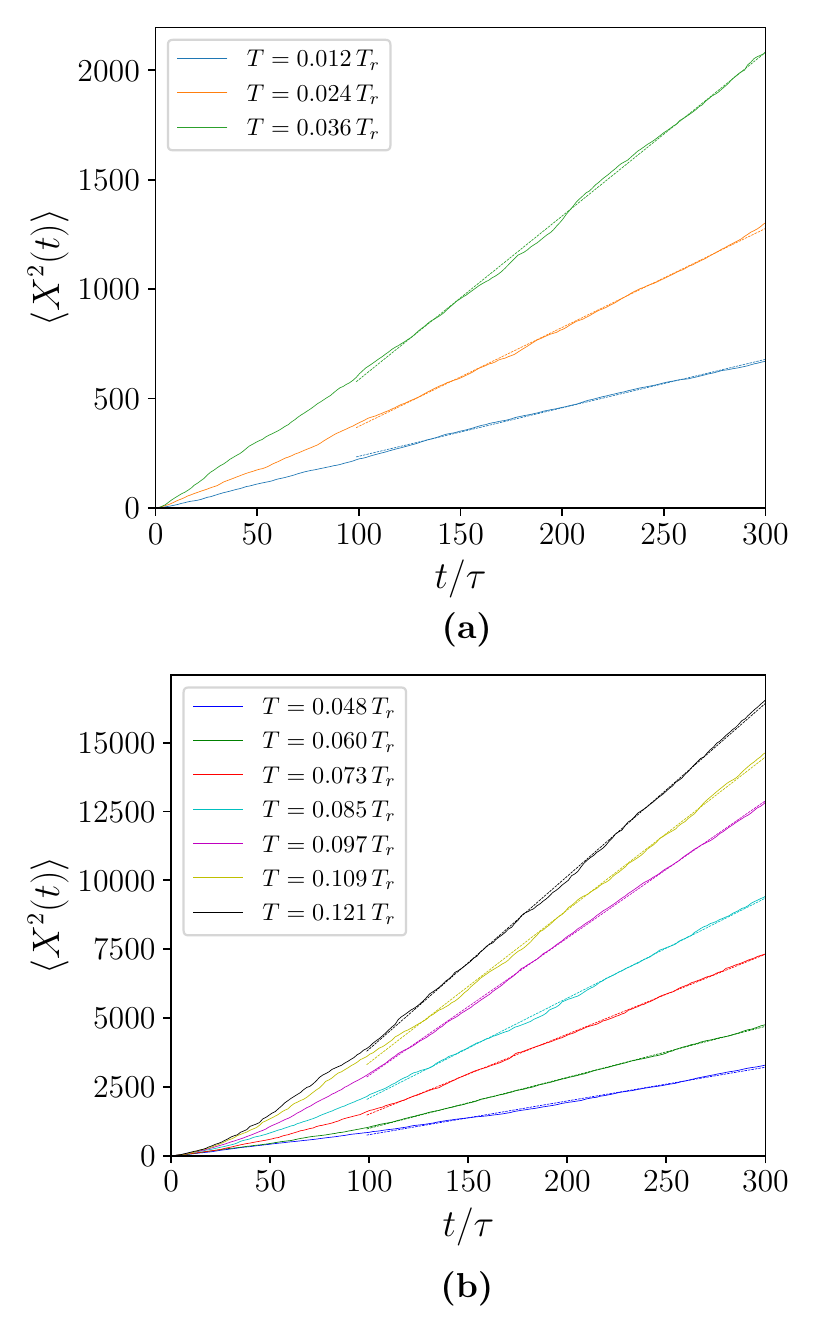}
    \caption{\label{fig:6} Polaron mean-square-displacement as a function of time in (a) the lowest temperature regime, $T \ll T_r$, and (b) the intermediate temperature regime, $T < T_r$. In both cases the dashed lines are linear fits to the data (in solid) taken after an equilibration time of $100\tau$.}
\end{figure}

The displacement of the center of mass positions leads to a net diffusion of the polaron. This is visualised via the mean-squared displacement plots in \cref{fig:6}. Keeping track of the center of mass positions across periodic boundaries is achieved by centering the chain indices around the center of mass position at each time step, and calculating displacements relative to this position. As displacements between consecutive discrete time steps are expected to be small, this approach avoids artificial and unphysically large jumps that would result from the center of mass crossing the periodic boundary in real space. The mean-squared displacement, $\langle X^{2}(t) \rangle$, is calculated as
\begin{align}
 \langle X^{2}(t) \rangle = \sum_{k=1}^{N_{t} = t/\Delta t} \left\{\bar{n}\left[k\Delta t\right] - \bar{n}\left[(k-1)\Delta t\right]\right\}^2,
\end{align}
where $\Delta t$ is chosen so that small fluctuations around the same physical position on the chain are averaged out. $\Delta t \sim \tau$ was found to give good results. Across all temperature values it is diffusive behavior, i.e., a linear change in the mean-squared displacement over time is achieved after an initial equilibration period of $\sim 100 \tau$.

\begin{figure}[h!]
    \centering
    \includegraphics[width=0.9\myfigwidth]{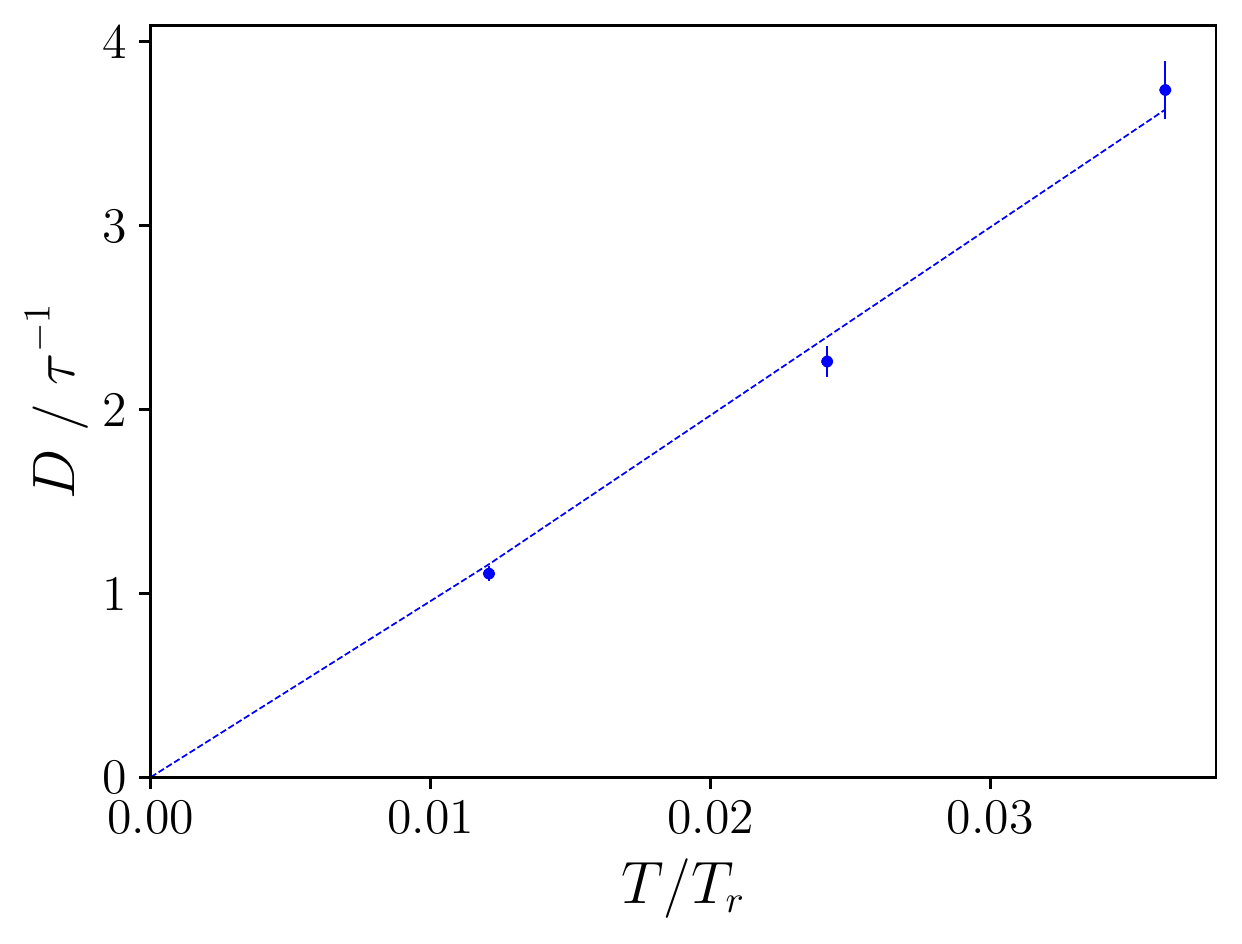}
    \caption{\label{fig:7} Diffusion constant, $D$, as a function of temperature in the lowest temperature regime, $T \ll T_r$, indicating an activationless process.}
\end{figure}

The gradient of the diffusion plots, taken after the initial equilibration, gives an estimate of the diffusion constant, $D$, as
\begin{align}
D = \frac{1}{2d}\lim_{t\to\infty} \dd{\langle X^{2}(t) \rangle}{t},
\end{align}
where $d$ is the dimensionality of the system, so $d=1$ in the case of quasi-1D systems such as the model under investigation. The diffusion constant is expected to be proportional to the temperature for activationless processes, and this is what is observed in the low temperature regime, as shown in \cref{fig:7}.
These results are in agreement with previous work.\cite{Albu13a,Poole16,Tozer15,Binder20a}

\subsection{Intermediate temperature polaronic hopping}

\begin{figure}[h!]
    \centering
    \includegraphics[width=0.9\myfigwidth]{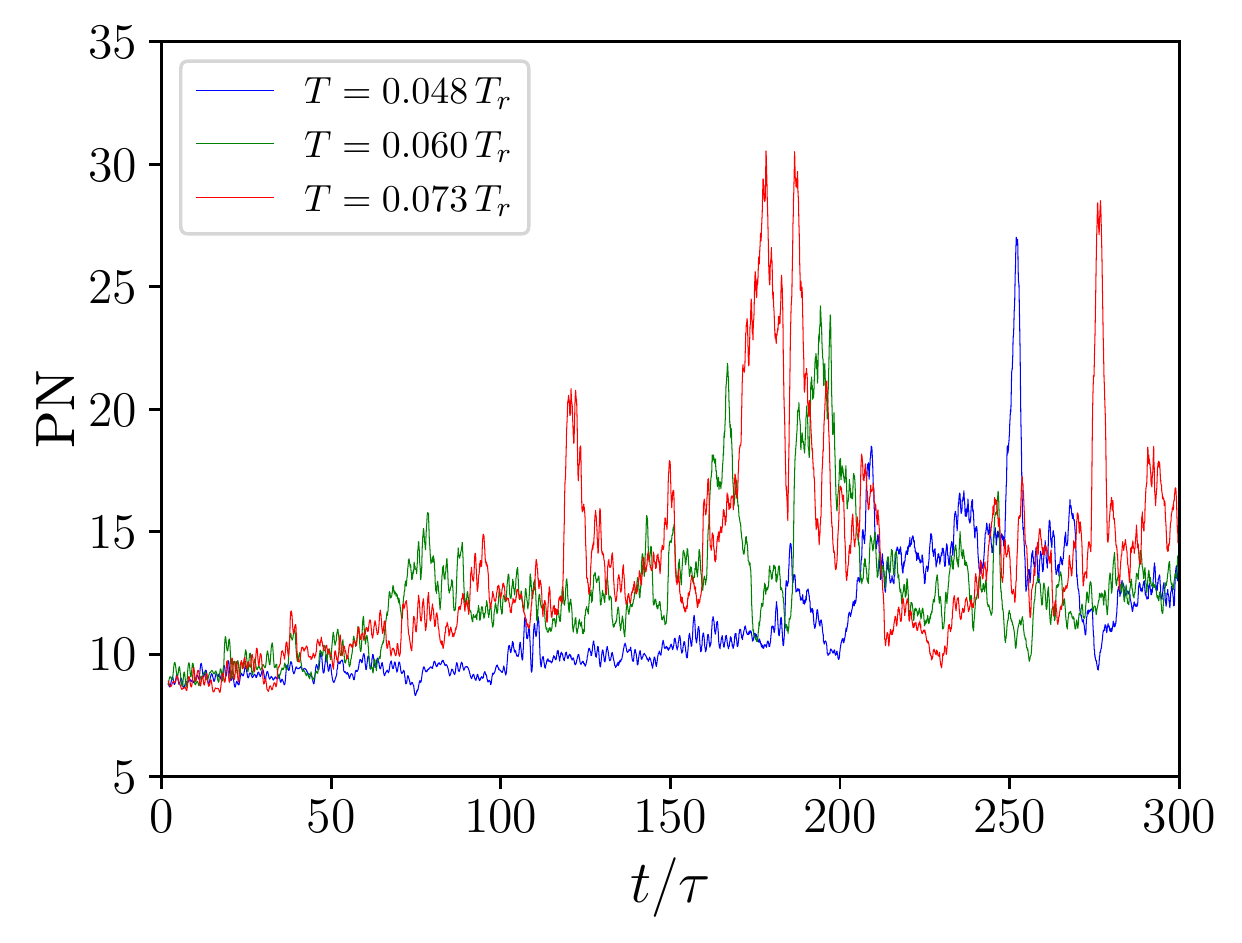}
    \caption{\label{fig:8} Participation number (i.e., localization length) of the evolving state, $|\Psi(t)\rangle$, as a function of time at different temperatures. Note the increase in PN when $T = 0.073\, T_r$ at $t \sim 170  \tau$ corresponding to the Landau-Zener transition shown in \cref{fig:11}. The curves are smoothed for better legibility.}
\end{figure}
\begin{figure}[h!]
    \centering
    \includegraphics[width=0.9\myfigwidth]{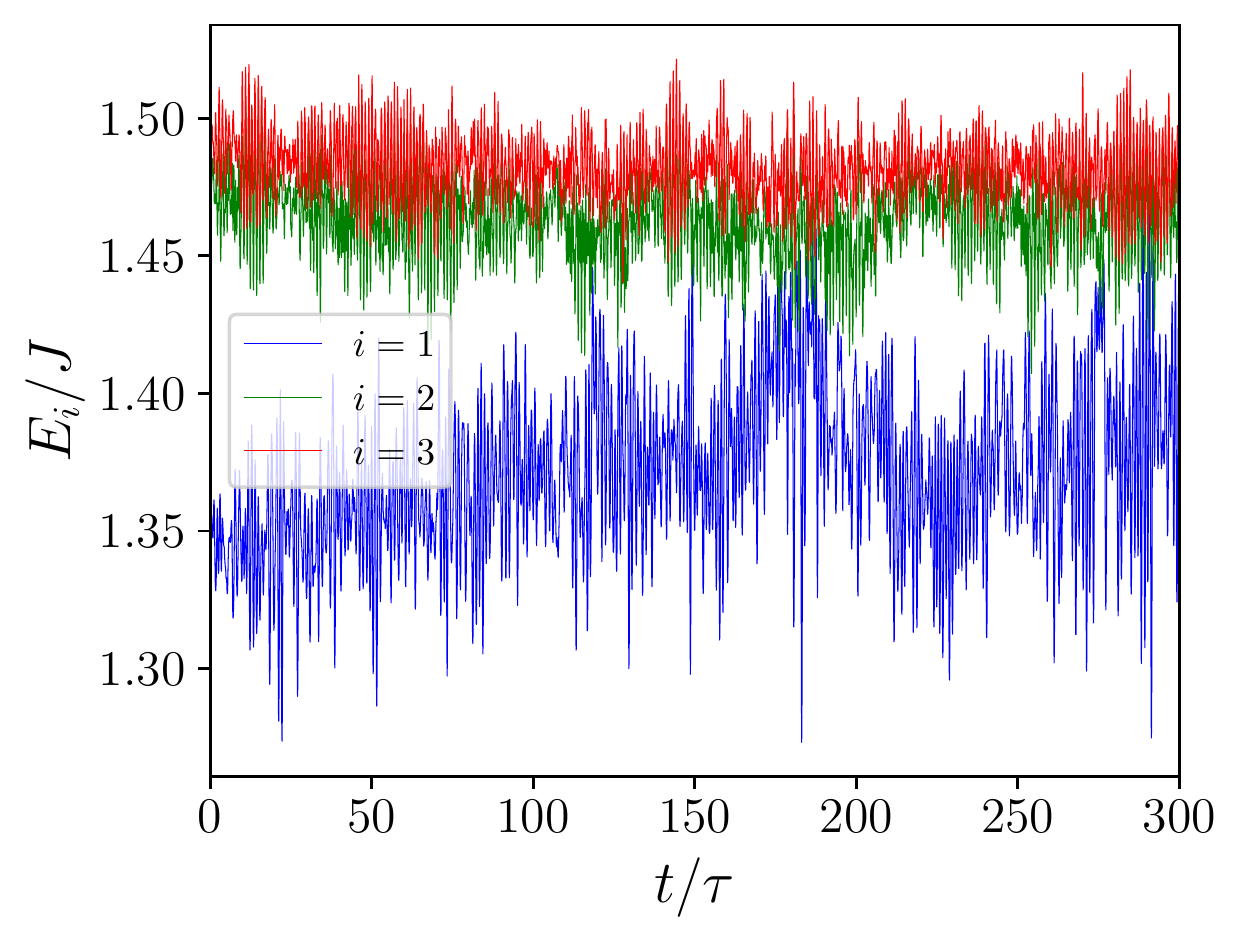}
    \caption{\label{fig:9} Energies of the lowest instantaneous eigenstates as a function of time at $T = 0.073\, T_r$. Note the avoided crossing at $t \sim 170  \tau$ corresponding to the Landau-Zener transition shown in \cref{fig:11}.}
\end{figure}
\begin{figure}[h!]
    \centering
    \includegraphics[width=0.9\myfigwidth]{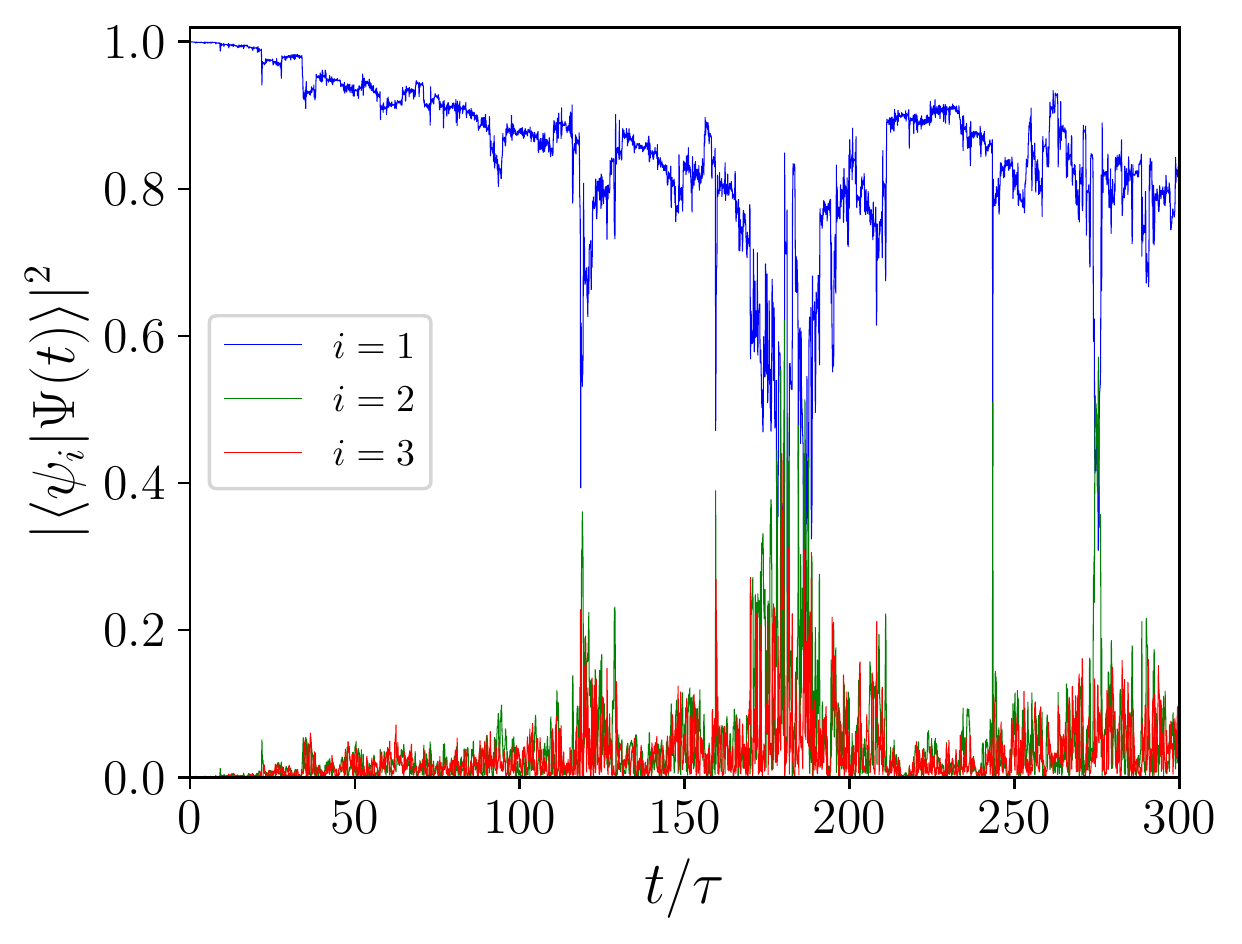}
    \caption{\label{fig:10} Probabilities that the evolving state, $|\Psi\rangle$, occupies an instantaneous adiabatic state, $|\psi_i\rangle$, as a function of time at $T = 0.073\,T_r$. Note that the probabilities of occupying $|\psi_1\rangle$ and $|\psi_2\rangle$ become almost equal at $t \sim 170  \tau$ corresponding to the Landau-Zener transition shown in \cref{fig:11}.}
\end{figure}

\begin{figure*}[h!]
    \centering
    \includegraphics{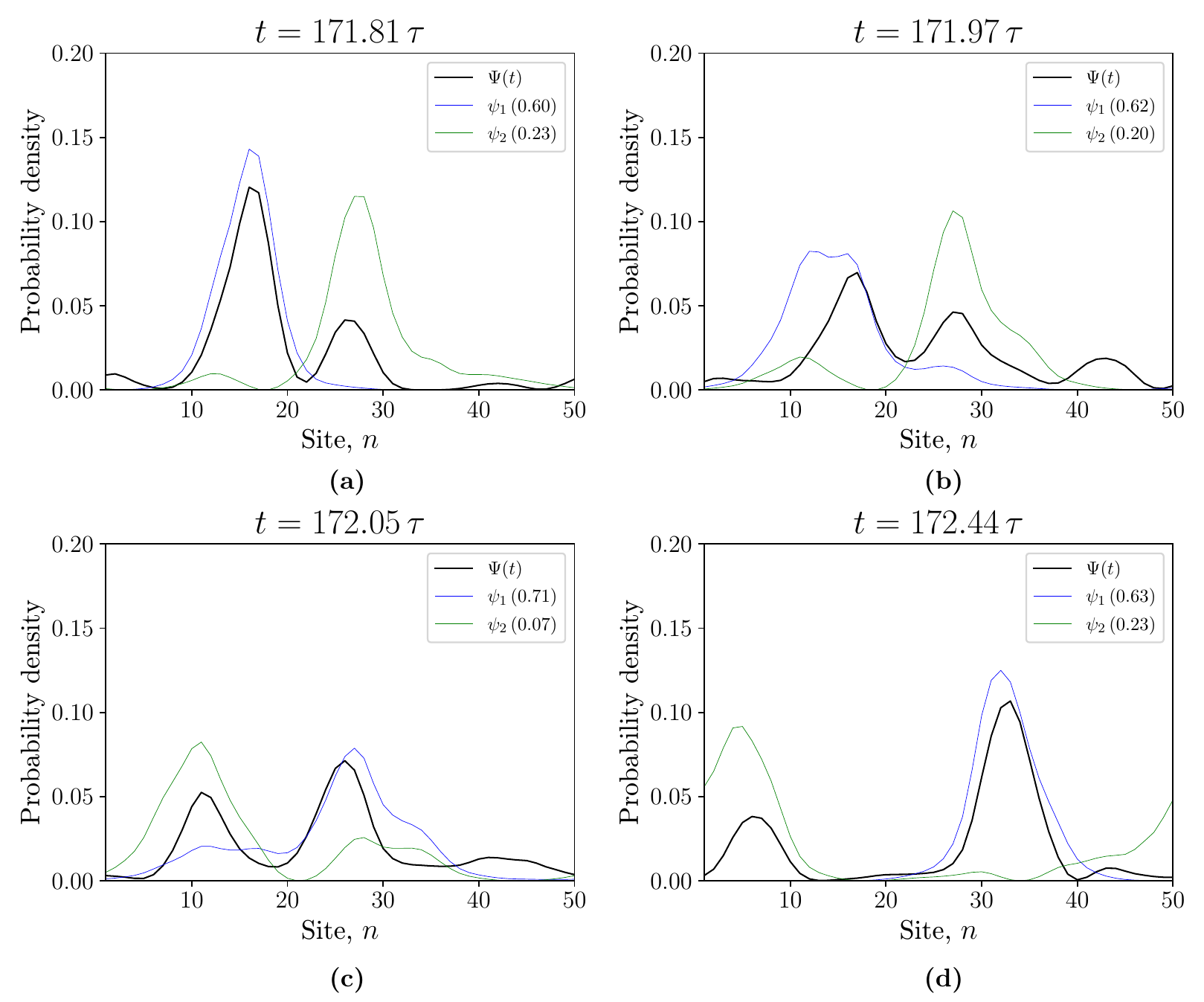}
    \caption{\label{fig:11} Evolution of $|\Psi\rangle$ (black) and the two lowest adiabatic states ($|\psi_1\rangle$ (blue) and $|\psi_2\rangle$ (green)) during a Landau-Zener transition when $t \sim 170 \tau$ at $T = 0.073\, T_r$. The curves represent the probability densities. Note that $|\Psi_n|^2 = \rho_n \sim \Delta\tilde{\phi}_n^{\textrm{eq}}  \times (-1)^n$. The occupation probabilities of $|\psi_1\rangle$ and $|\psi_2\rangle$ are shown in the legends, showing that during the transition the system remains predominately in the lower adiabatic state. As described in the text, the transition occurs between panels (b) and (c).}
\end{figure*}

\begin{figure}[h!]
    \centering
    \includegraphics[width=0.9\myfigwidth]{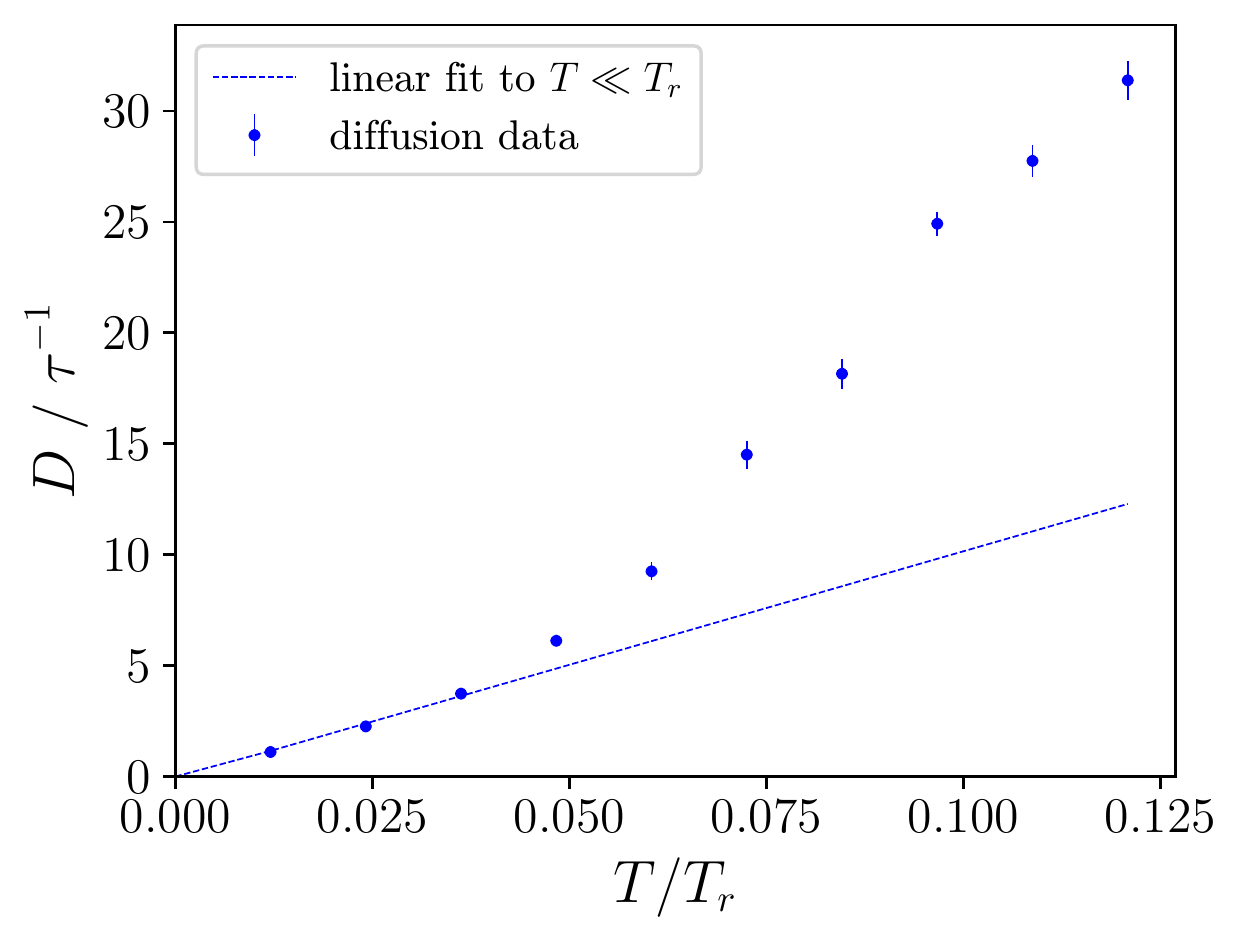}
    \caption{\label{fig:12} Diffusion constant, $D$, as a function temperature, indicating a crossover to activated hopping processes at higher temperatures.}
\end{figure}
\begin{figure}[h!]
    \centering
    \includegraphics[width=0.9\myfigwidth]{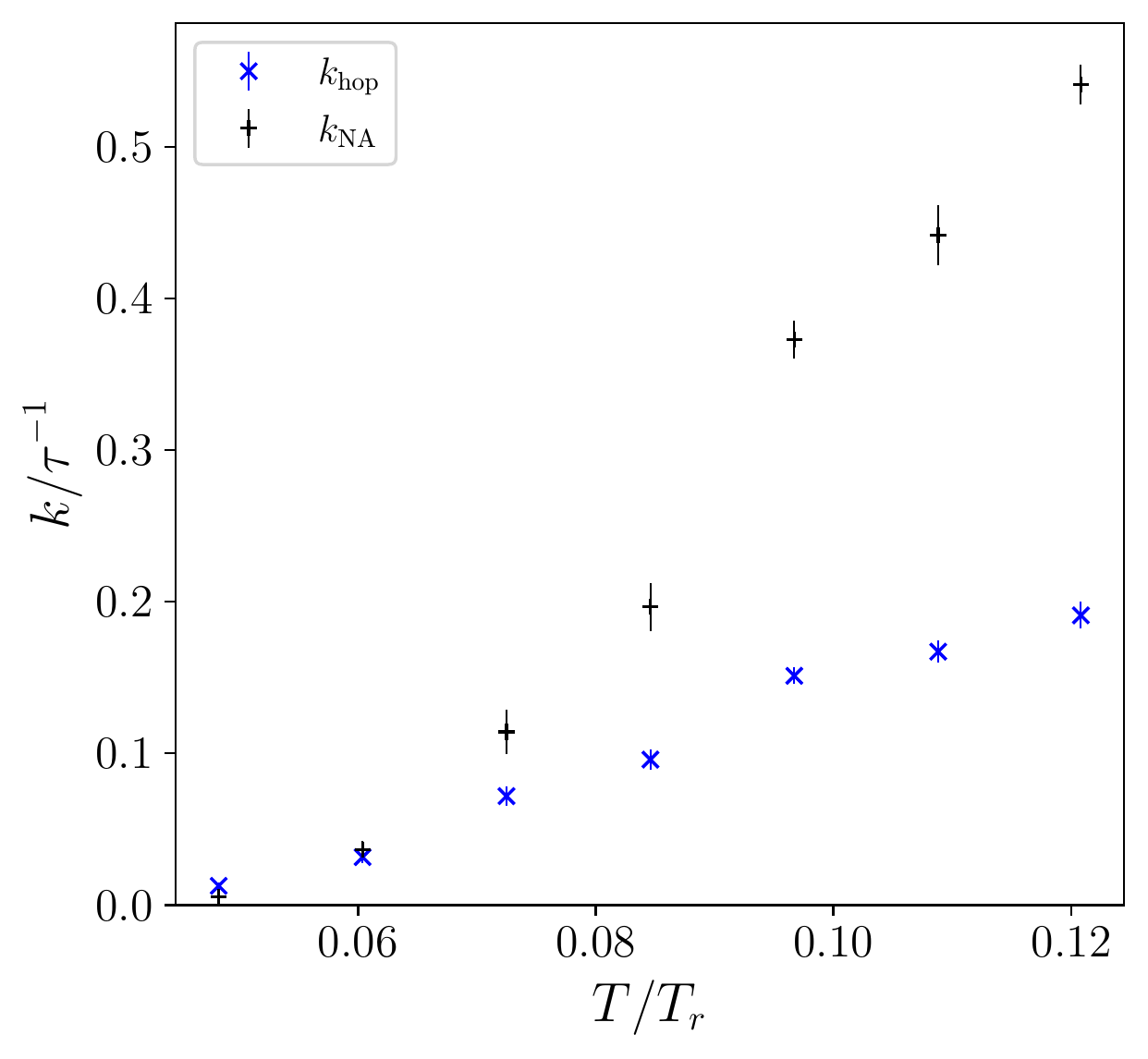}
    \caption{\label{fig:13} Hopping transition rates calculated via \cref{Eq:22} using a mean displacement, $\langle \Delta r \rangle = 10\, \mathrm{ sites}$.
The rate for nonadiabatic processes, $k_{\textrm{NA}}$, is defined as the rate at which the probability of the system occupying the lowest adiabatic state $\ket{\psi_1}$ becomes less than $0.5$.}
\end{figure}

As the temperature is increased the larger torsional fluctuations lead to a characteristically different type of dynamics. The evolving state, $|\Psi(t)\rangle$, is no longer localized at all times (as shown by the PN in \cref{fig:8}), the energies of the instantaneous adiabatic eigenstates exhibit (avoided) crossovers (as shown in \cref{fig:9}), and the probability of occupying higher-lying eigenstates increases considerably at various times (as shown in \cref{fig:10}).

At the points of energetic (avoided) crossings, Landau-Zener transitions\cite{Landau1932a,Landau1932b,Zener1932,Zener1933,Nitzan2006book} are observed. An example of this is shown in a series of snapshots in \cref{fig:11} taken at $t \sim 170 \tau$ (see also \cref{fig:8,fig:9,fig:10}).
\Cref{fig:11} shows the probability densities of the evolving state, $|\Psi(t)\rangle$, and the two lowest instantaneous adiabatic eigenstates, $|\psi_1\rangle$ and $|\psi_2\rangle$. \Cref{fig:11}(a) shows that $|\Psi(t)\rangle$ is a linear combination of $|\psi_1\rangle$ and $|\psi_2\rangle$, but predominately composed of $|\psi_1\rangle$. At this time the adiabatic states $|\psi_1\rangle$ and $|\psi_2\rangle$ are predominately composed of separate quasidiabatic states localized at $n \sim 15$ and $n \sim 28$, respectively. Passing through the transition [(b) and (c)], we observe that the adiabatic states become linear combinations of the diabatic states, and similarly for $|\Psi(t)\rangle$. The transition occurs between (b) and (c), as in (b) $|\psi_1\rangle$ is predominately the left-hand diabatic state, while in (c) it is predominately the right-hand diabatic state. In (d) the adiabatic states are again predominately composed of separate quasidiabatic states localized at $n \sim 5$ and $n \sim 33$, respectively. Throughout, $|\Psi(t)\rangle$ is predominately composed of $|\psi_1\rangle$, i.e., it remains on the lower adiabatic energy surface. This avoided-crossing event thus corresponds to a transition between quasidiabatic states, or a `hop' between polaron states.

This transition occurs because the thermally induced fluctuations in the monomer conformations, $\phi$, cause the polaron to be less stable than its zero-temperature configuration, thus raising its energy (as shown by the blue curve at $t = 170 \tau$ in \cref{fig:9}) and to increase its size (as shown by the blue curves in \hyperref[fig:11]{Figs. 11}(a) and \hyperref[fig:11]{11}(b). Simultaneously, the conformational fluctuations cause the energy of a neighboring diabatic state to decrease, causing a Landau-Zener type transition.
Indeed, since the charge density, $\rho_n$, (given by the black curve) and the angular displacements, $\Delta\tilde{\phi}_n^{\textrm{eq}}$, satisfy \cref{Eq:10}, \cref{fig:11}(b) and \hyperref[fig:11]{11}(c) show that angular displacements of each diabatic state are essentially identical, thus exhibiting a classic Marcus-type transition.

The transfer of population probability between two quasidiabatic states that are localized on different sections of the polymer chain leads to a net displacement of the center of mass on a faster timescale than that of the crawling diffusion described in the low temperature regime in the previous section.
This is observed in the diffusion constants calculated from the $\langle X^{2}(t) \rangle$ plots, where the diffusion parameters deviate from the linear $T$ behavior observed at low temperatures. This is shown in \cref{fig:12}, where the linear fit to the low temperature data points can be seen to underestimate the diffusion in the higher temperature regime due to the additional displacement from hopping transport.

To quantify this behavior, hopping transition rates were calculated by assuming that the deviation from the linear temperature dependence of $D(T)$ (shown in \cref{fig:12}) is attributed to hopping transport. Assuming a mean displacement $\langle \Delta r \rangle =10$ when such a transition occurs (as this value is consistent with a PN = 10 as well as to the hop distance illustrated in \cref{fig:11}), the transition rate for the process is calculated as
\begin{gather}\label{Eq:22}
k_{\mathrm{hop}} \propto \frac{D_{\mathrm{hop}}}{\langle \Delta r \rangle^2},
\end{gather}
with the results plotted in \cref{fig:13}.


\section{Discussion and Concluding Remarks}\label{sec:conclusions}

This paper has described our simulations of charge-polaron transport in conjugated polymers, valid in the large-polaron limit. However, as the model is general it equally applies to exciton-polaron (i.e., energy) transport. For example, \cref{eq:Ham} is derived for Frenkel excitons in Ref.\cite{Barford18}, with the exciton transfer integral $J_n^0 = J_{DD} + J_{SE} \cos^2 \theta_n^0$,
where $J_{DD}$ and $J_{SE}$ are the through-space and through-bond transfer integrals, respectively, and the exciton-roton coupling constant is
$A_n = {J_{SE}\sin 2 \theta_n^0}/{(\hbar \omega K)^{1/2}}$.

Charge or exciton coupling to the monomer rotations localizes the particle into a Landau polaron, while the
thermal fluctuations of the monomers cause polaron dynamics.
The emergent low-energy scale of our model is the polaron reorganization energy, $E_r$, which corresponds to a reorganization temperature $T_r \sim 400 \,\si{\kelvin}$ for charges and $T_r \sim 1500 \,\si{\kelvin}$ for excitons.\cite{Tozer15}
We investigated two types of dynamics -- both relevant for temperatures $T < T_r = E_r/k_B$.
As found before,\cite{Albu13a, Poole16,Tozer15,Binder20a} in the lower temperature regime the system remains in the same quasidiabatic state, corresponding to activationless polaron diffusion as the polaron crawls stochastically along the chain, with $D(T) \propto T$. As the temperature is raised, however, there is a cross-over to an additional activated transfer process between diabatic states. We showed that these hopping processes exhibit Landau-Zener dynamics whose origin is the fluctuation of torsional angles causing a temporary degeneracy of the energy levels of neighboring diabatic states, as described in the Marcus theory of charge transfer.

Our hybrid quantum-classical simulations treat the charge or exciton dynamics fully quantum mechanically. However, as explained in the Introduction, to simulate large systems over long times, we model the nuclear (i.e., monomeric) degrees of freedom via the Ehrenfest approximation and the solvent via the Langevin equation.
The assumptions of the Ehrenfest approximation, namely that the nuclear degrees of freedom are treated classically and the total wave function is a product of the electronic and nuclear wave functions, is known to lead to a number of incorrect theoretical predictions.\cite{Horsfield06,Nelson20} In particular,
a simple product wave function implies that the nuclei move in a mean potential determined by the electrons. This means that a splitting of the nuclear wave packet when passing through a conical intersection or an avoided crossing does not occur. In addition, there is an incorrect description of energy transfer between the electronic and nuclear degrees of freedom, meaning that detailed balance is not necessarily satisfied.

Burghardt \textit{et al.}\cite{Hegger20} showed that an Ehrenfest simulation was qualitatively consistent with a fully quantum mechanical approach on small systems, arguing that this is because the dynamics was quasiadiabatic. We have attempted to determine whether or not our hopping transitions are quasiadiabatic by estimating the rate for nonadiabatic events, $k_{\textrm{NA}}$. This is defined as the rate at which the probability of the system described by $\ket{\Psi(t)}$ occupying the lowest adiabatic state $\ket{\psi_1}$ becomes less than $0.5$. This is shown in \cref{fig:13}. For the lowest temperature data points we see that $k_{\textrm{NA}} \approx k_{\textrm{hop}}$, but that $k_{\textrm{NA}}$ exceeds $k_{\textrm{hop}}$ as the temperature increases. Assuming that $k_{\textrm{hop}}$ is a measure of the rate for quasiadiabatic events (i.e., Landau-Zener transitions that predominately stay on the lower surface and hence cause a diabatic transition), these results imply that our observed Landau-Zener transitions are never happening under a fully quasiadiabatic regime, and indeed nonadiabatic events dominate as the temperature increases.
This observation does not necessarily imply that the Ehrenfest approximation is failing and that there is anomalous heating, however, because as shown in \cref{fig:6} after the equilibration period of $\sim 100 \tau$ the gradient of the mean-square-displacement versus time is constant for a fixed temperature.

Our simulations become difficult to interpret as $T$ approaches $T_r$, so we have not been able to fit an Arrhenius or Marcus expression to our rate constant in the activated regime. However, our simulations do show conclusive evidence of Landau-Zener type transitions driven by thermal fluctuations in a realistic model.

Qualitatively, we expect that at $T \sim T_r$, because of the large density of states of higher-lying eigenstates, the polaron becomes unbound. Then the charge or exciton exhibits quasiband-like transport via the quasiextended (albeit Anderson localized) particle wave functions.\cite{Malyshev01a,Malyshev01b,Makhov10}

Another assumption of treating the oscillators and bath classically is that they obey the principle of equipartition. A quantum mechanical treatment of both would freeze-out degrees of freedom at low temperatures, thus suppressing monomer rotations. In the quantum limit, although polaron crawling will still occur, the diffusion constant will be smaller than its classical prediction.

In principle, our model could be adapted by including surface hopping (see e.g., Ref.\cite{Wang15}). Alternatively, within the TNT Library we could quantize the harmonic oscillators and introduce Lindblad jump operators to model relaxation through conical intersections, thus describing both the high and low temperature limits.\cite{Mannouch2018} However, both schemes require considerably more computational effort.

Our results have qualitative relevance for other theory\cite{Garg1985,Takahashi2017} and to experimental observations of charge dynamics in conjugated polymers. In particular, our demonstration of Landau-Zener type charge hopping provides qualitative numerical justification for the fluctuating-bridges mechanism of charge transfer\cite{Skourtis2003,Troisi2003,Fornari14}. This mechanism leads to a corrected Marcus theory of unimolecular charge transport which was employed by Berencei \textit{et al.} to simulate charge transport in conjugated polymer systems.\cite{Berencei19}

\begin{table}
\small\centering
{\renewcommand{\arraystretch}{1.2}
\begin{tabular}{|p{0.45\linewidth}|p{0.45\linewidth}|}
\hline
Angular  & Linear \\
\hline
Angular displacement, $\Delta\phi$ & Linear displacement, $u$ \\
Angular momentum, $L$ & Linear momentum, $p$ \\
Torque, $\Gamma = -\partial \langle \hat{H}\rangle/\partial \Delta \phi$  & Force, $f = -\partial \langle \hat{H}\rangle/\partial u$ \\
Inertia, $I$ & Mass, $m$ \\
$K$ (energy rad$^{-2}$) & $K$ (energy length$^{-2}$) \\
\hline
\end{tabular}}
\caption{The equivalence of \cref{eq:Ham} for linear nuclear degrees of freedom.}
\label{ta:1}
\end{table}

In this paper we have motivated charge and exciton polaron transport in 1D systems by considering intrachain motion in conjugated polymers driven by the thermal angular fluctuations of the monomers. However, our model is rather general. As shown in \cref{ta:1}, by replacing the angular variables by linear variables, \cref{eq:Ham} becomes essentially equivalent to the Su-Schrieffer-Heeger model.\cite{Su79} Charge mobility was investigated in 1D molecular stacks using this model by Troisi and Orlandi\cite{Troisi06} and by Wang \emph{et al.}\cite{Wang15} At high temperatures both sets of authors observed quasiband-like behavior, where the mobility $\mu \sim T^{-{\alpha}}$. We predict that this transport occurs at temperatures $T \sim T_r = E_r/k_B$.
At lower temperatures Wang \emph{et al.}\cite{Wang15} observed Marcus-hopping transport, although quasidiabatic polaron crawling was not observed by them at even lower temperatures. Our future goal is to model all three regimes within the same simulation.

\begin{appendix}

\section{Derivation of \cref{eq:Ham} for Charges}\label{App:1}

\Cref{eq:Ham} describes a coarse-grained charge-roton model for the PPP polymer chain. As shown in our earlier work,\cite{Berencei19} the model can be derived starting from the atomic H\"uckel model. In the atomic basis of $p_z$-orbitals, the Hamiltonian can be written as
\begin{gather}
    \hat{H}^{\mathrm{atomic}}_{\mathrm{H\ddot{u}ckel}} = \sum_{j} \beta_{j}\left( \creat{j}\annil{j+1} + \creat{j+1}\annil{j} \right),
\end{gather}
where $\beta_j$ is the transfer integral between nearest neighbor atomic orbitals, and $\creat{j}$ $(\annil{j})$ creates (destroys) an electron in orbital $j$. There are no on-site terms as the $p_z$ orbitals are degenerate. Coarse-graining the polymer into a chain of `sites' (illustrated in \cref{fig:1}) reduces the degrees of freedom in the model. In the coarse-grained model, the atomic orbital basis $\{ \vert j \rangle \}$ is replaced by the `site' basis $\{ \vert n \rangle \}$, leading to redefined creation (destruction) operators:
\begin{gather}
\vert j \rangle = \creat{j}\vert 0 \rangle \quad \Rightarrow \quad \vert n \rangle = \creatcg{n} \vert 0 \rangle,
\end{gather}
resulting in the coarse-grained H\"uckel model,
\begin{gather}
    \hat{H}^{\mathrm{cg}}_{\mathrm{H\ddot{u}ckel}} = \sum_{n}\epsilon\creatcg{n}\annilcg{n} + \sum_{n} J_{n}\hat{T}_{n,n+1},
\end{gather}
where $\epsilon$ represents the uniform monomeric LUMO energy, and
\begin{gather}
\hat{T}_{n,n+1} = \left( \creatcg{n}\annilcg{n+1} + \creatcg{n+1}\annilcg{n} \right)
\end{gather}
is the symmetric jump operator in the coarse-grained basis.
The sum over $n$ is over `sites', i.e., the coarse-grained monomers.

The transfer term is
\begin{gather}
    \label{eq:App_J}J_n = \frac{1}{3}\beta_{0}\cos{\theta_n},
\end{gather}
where $\beta_{0}$ is the H\"uckel resonance integral between parallel $p_z$ orbitals. The factor of $1/3$ arises as the product of orbital coefficients on the carbon atoms linking the two sites, as shown in \cref{fig:1}. $J_n$ is expressed as a function of the torsional angle, $\theta_{n}$, across bond $n$, where $\theta_{n}$ is defined as the difference between the rotational angles on neighboring monomers,
\begin{gather}
\theta_{n} = \phi_{n+1} - \phi_{n}.
\end{gather}

Adding the rotational potential and kinetic energy terms, the coarse-grained charge-roton Hamiltonian for the PPP model system reads,\cite{Berencei19}
\begin{align}
\label{eq:Ham_er_full} \nonumber \hat{H} = &\sum^{N}_{n=1} \left(\epsilon\creatcg{n}\annilcg{n} + J_{n}\hat{T}_{n,n+1} \right)\\ &+ \sum^{N}_{n=1}\left(\frac{K}{2} (\Delta \phi_{n})^2 + \frac{L^2_{n}}{2I} \right).
\end{align}

Assuming small angular displacements from equilibrium in the neutral state, $\phi^{0}_{n}$,
\begin{gather}
\label{eq:J_linear} J_{n} \simeq \beta_{0} \cos{\theta^{0}_{n}}/3 - \beta_{0} \sin{\theta^{0}_{n}}\times(\Delta \phi_{n+1} - \Delta\phi_{n})/3.
\end{gather}

Finally, introducing dimensionless variables
\begin{gather}
\tilde{\phi} = (K/\hbar\omega)^{1/2}\phi, \\
\tilde{L} = (\omega/\hbar K)^{1/2} L,
\end{gather}
using \cref{eq:J_linear} and ignoring the uniform on-site energy terms, \cref{eq:Ham_er_full} becomes
\begin{align}
\label{eq:Ham_OBC} \nonumber \hat{H} = &\sum_{n=1}^{N-1}\left[ J_n^0 - \hbar\omega A_n (\Delta \tilde{\phi}_{n+1} - \Delta\tilde{\phi}_{n})\right]\hat{T}_{n,n+1}  \\ &+ \frac{\hbar\omega}{2} \sum_{n=1}^{N} \left( (\Delta \tilde{\phi}_{n})^{2} + \tilde{L}_{n}^{2} \right),
\end{align}
where
\begin{equation}
J_n^0 = \beta_{0} \cos \theta_n^0/3
\end{equation}
and
\begin{gather}
A_n = \frac{\beta_{0}\sin{\theta^{0}_{n}}}{3(\hbar\omega K)^{1/2}}
\end{gather}
is the dimensionless charge-roton coupling constant.

\end{appendix}

\begin{acknowledgments}
This work was performed using the Tensor Network Theory Library, Beta Version 1.2.1 (2016), S. Al-Assam, S. R. Clark, D. Jaksch, and the TNT Development team.\cite{Al_Assam_2017}
L.B. is supported by the EPSRC Centre for Doctoral Training, Theory and Modelling in Chemical Sciences (Grant No. EP/L015722/1). S.R.C. gratefully acknowledges support from the EPSRC under grant EP/T028424/1.
\end{acknowledgments}

\section*{DATA AVAILABILITY}

The data that support the findings of this study are available from the corresponding author upon reasonable request.

\bibliography{references}

\clearpage

\section*{Supplemental Material}
\label{sec:supp}

\setcounter{section}{0}
\renewcommand{\thesection}{S-\Roman{section}}
\setcounter{equation}{0}
\renewcommand{\theequation}{S\arabic{equation}}
\setcounter{figure}{0}
\renewcommand{\thefigure}{S\arabic{figure}}

\section{Periodic Boundary Conditions in TNT}
\begin{figure}[h!]
\centering
\includegraphics[width=0.4\textwidth]{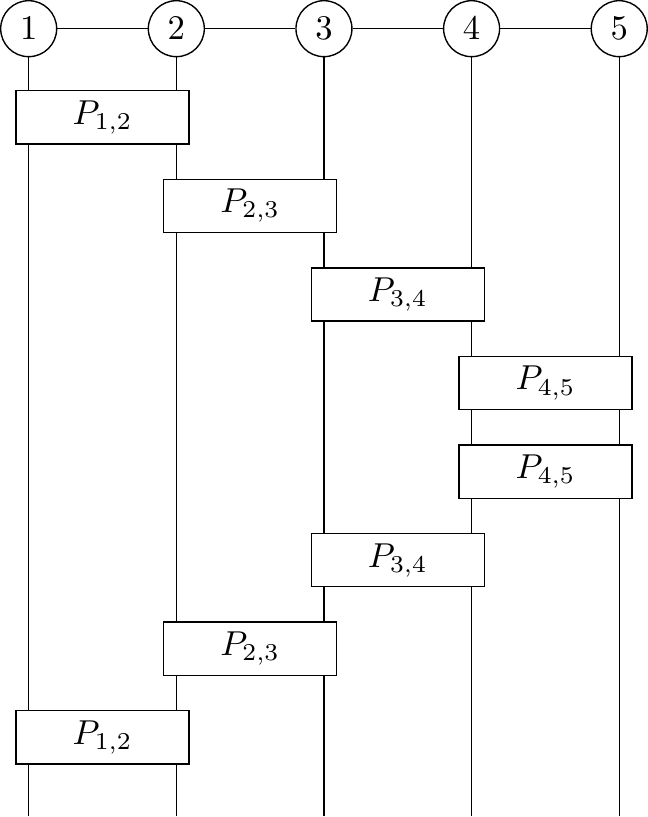}
\caption{Trotter time evolution in OBC system.}
\label{fig:staircase_OBC}
\end{figure}

In order to overcome finite-size effects and add the possibility of appropriately simulating macrocyclic materials, periodic boundary conditions (PBC) needed to be implemented in the TNT library,\cite{Al_Assam_2017} specifically in the time-evolving block decimation (TEBD)\cite{Vidal2003,Vidal2004} scheme. As mentioned in the main paper, the time-dependent Schr\"odinger equation is solved via
\begin{gather}
  \vert \Psi_{\mathrm{MPS}}(t+\delta t) \rangle = \exp(- \compi\hat{H}\delta t/\hbar)\vert \Psi_{\mathrm{MPS}}(t) \rangle
\end{gather}
In the TEBD scheme, the Hamiltonian is first rewritten as a sum of two-site terms:
\begin{gather}
\hat{H} = \sum_{n=1}^{L-1} \hat{H}_{n,n+1},
\end{gather}
where each $ \hat{H}_{n,n+1}$ only contains terms for sites $n$ and $n+1$. This form of the Hamiltonian allows for the numerically efficient application of the time propagator using a second-order Suzuki-Trotter-decomposition.\cite{Trotter1959,Suzuki1976,DeRaedt1983}
\begin{gather}
\label{eq:Trotter}\expe^{-\compi\hat{H}\delta t} \approx \expe^{-\frac{\compi}{2}\hat{H}_{1,2}\delta t}\expe^{-\frac{\compi}{2}\hat{H}_{2,3}\delta t}\cdots\expe^{-\frac{\compi}{2}\hat{H}_{2,3}\delta t}\expe^{-\frac{\compi}{2}\hat{H}_{1,2}\delta t}+\mathcal{O}(\delta t^3)
\end{gather}
The time-evolution `staircase' in TEBD mirrors the structure of the Suzuki-Trotter decomposition, and for and open boundary (OBC) system, can be represented diagrammatically as shown in \cref{fig:staircase_OBC}. The numbered circles represent each site in the MPS wave function, and each $P_{i,i+1}$ `gate' represents a half-timestep propagation in accordance with the Trotter decomposed time evolution.

\begin{figure}[H]
\centering
\includegraphics[width=0.8\textwidth]{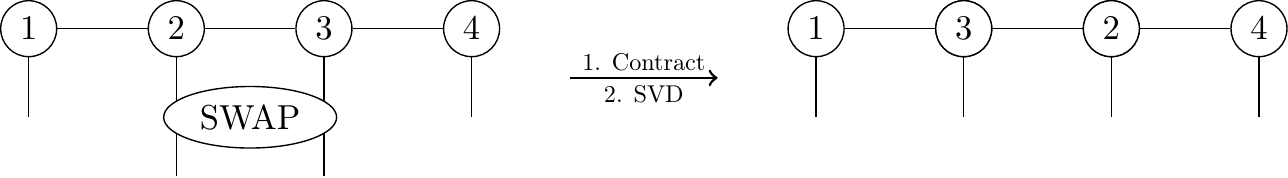}
\caption{Swap gate acting on chain of sites.}
\label{fig:swap}
\end{figure}

\begin{figure}[H]
\centering
\includegraphics[width=0.4\textwidth]{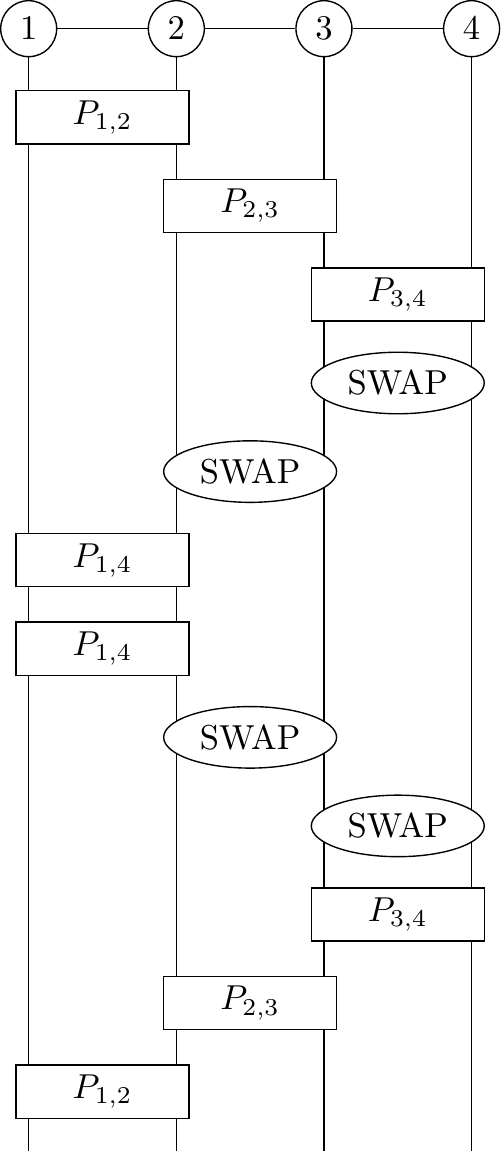}
\caption{Time evolution in PBC system. The periodic, end-site propagator is denoted as $P_{1,4}$.}
\label{fig:staircase_PBC}
\end{figure}

Periodicity in the system is introduced through additional interactions added to the system considered, which link the end sites of the chain of sites. As presented above, for numerical efficiency, the time propagation is implemented through a sequence of nearest-neighbor interactions. In order to be able to extend this algorithm to periodic systems, when end site interactions are applied, these sites are required to be in neighboring positions in our MPS wave function. Achieving this is possible through the application of a series of `swap gates', which have the role of swapping neighboring physical indices along the 1D system. For example, for two sites spanned by a basis of four states the swap gate takes the following form:
\[ \makeatletter\setlength\BA@colsep{6pt}\makeatother
\mathrm{SWAP} = \begin{blockarray}{ccccc}
\ket{00} & \ket{01} & \ket{10} & \ket{11} \\
\begin{block}{(cccc)c}
  1 & 0 & 0 & 0 & \ket{00} \\
  0 & 0 & 1 & 0 & \ket{01} \\
  0 & 1 & 0 & 0 & \ket{10} \\
  0 & 0 & 0 & 1 & \ket{11} \\
\end{block}
\end{blockarray}.
 \]
 The effect of applying this gate to a chain of sites is shown diagrammatically in \cref{fig:swap}.
The addition of a series of swap gates within the time propagation process is therefore required in order to follow the correct Trotter decomposition with the additional end site interaction term. This is shown diagrammatically in \cref{fig:staircase_PBC}.
The accuracy of the PBC implementation was tested against analytic solutions of simple Hamiltonians and was found to be accurate, but often requires a smaller integration time step. In addition, due to the additional operations required for the site swaps at each time step the simulations take twice the time compared to the corresponding OBC calculations.


\end{document}